\documentstyle[pre,aps,multicol,epsf]{revtex}

\begin{document}   

\tightenlines

\title{Bekki-Nozaki Amplitude Holes in Hydrothermal Nonlinear Waves} 
\author{Javier Burguete,$^{1,2}$ Hugues Chat\'e,$^1$ 
	Fran\c{c}ois Daviaud,$^1$ and Nathalie Mukolobwiez$^1$}
\address{$^1$CEA --- Service de Physique de l'Etat Condens\'e, 
Centre d'Etudes de Saclay, 91191 Gif-sur-Yvette, France\\
$^2$Departamento de F\'{\i}sica y Matem\'atica Aplicada, Facultad de Ciencias,
Universidad de Navarra, 31080 Pamplona, Spain}
\maketitle
\begin{abstract}
We present and analyze experimental results on the dynamics of hydrothermal
waves occuring in a laterally-heated fluid layer. We argue that the large-scale
modulations of the waves are governed by a one-dimensional complex 
Ginzburg-Landau equation (CGLE). 
We determine quantitatively all the coefficients of this amplitude equation
using the localized amplitude holes observed in the experiment, 
which we show to be
well described as Bekki-Nozaki hole solutions of the CGLE.
\end{abstract}

\centerline{{\em Phys. Rev. Lett.} {\bf 82} (1999) p. 3252-3255}


\begin{multicols}{2} 

\narrowtext 

The status and nature of the so-called amplitude equations which can
be derived in the vicinity of symmetry-breaking instabilities is now
well-established \cite{MCCPCH,AMP-EQ}. They are ``universal'' in so far as 
they essentially depend on the symmetries of the physical
system and of its bifurcated solutions, but also because they 
often remain valid, at least at a qualitative level, even far away
from the instability threshold \cite{AWAY,CROQ}. 
However, determining accurately the coefficients 
of the underlying relevant amplitude equation from experimental data remains
a difficult task, especially in these far-from-threshold regimes.

The complex Ginzburg-Landau equation (CGLE), which describes the large-scale
modulations of the bifurcated solutions near oscillatory instabilities,
is perhaps the most-studied amplitude equation \cite{MCCPCH}. 
This priviledged situation
is due to both its relevance to many experimental situations and to the 
variety of its dynamical behavior, in particular its spatiotemporal chaos
regimes. One of the landmarks of the CGLE is that it possesses
localized ``defect'' solutions. Even in one space dimension, where no
topological constraint exists, numerical simulations
of the CGLE  \cite{NONLIN,HECKE}
and analytical \cite{WVSPCH,CONTE} work have revealed
the existence and importance of various amplitude hole solutions, which 
can often be seen as the ``building blocks'' of the complex spatiotemporal
dynamics observed. In particular, 
the one-parameter family of traveling hole solutions
discovered by Bekki and Nozaki \cite{NB} 
has been shown to play an important
dynamical role in a large portion of parameter space including in 
regions where they are linearly unstable \cite{NONLIN}.
Similar objects have been identified in various experimental contexts
of a priori relevance, e.g. Rayleigh-B\'enard convection 
and  coupled wakes \cite{EXP-HOLE}.
However, to our knowledge, there is still no case where
a direct comparison with known solutions of the CGLE 
could be achieved.

In this Letter, we present a quantitative comparison of localized
amplitude holes observed in an experiment with hole solutions
of the CGLE, the relevant amplitude equation.
We use the observed holes to fully determine the coefficients of
the underlying CGLE. This provides  clear-cut
evidence of Bekki-Nozaki holes in an experimental context.
Our system is a long, straight, and narrow  convection cell
in which a thin fluid
layer with a free surface is subjected to a horizontal temperature gradient.
Hydrothermal nonlinear waves appear via a direct Hopf bifurcation,
indicating the relevance of the CGLE.
The spatiotemporal dynamics of the waves exhibits
localized amplitude holes. The basic scales of the equivalent CGLE 
are determined using the regular part of the wave trains.
Data collected in the vicinity of amplitude holes
show that they have the structure of Bekki-Nozaki solutions. 
This also provides 
estimates of the remaining coefficients of the CGLE, an approach which, 
we argue, could be efficient in other experimental contexts.
Finally, the overall consistency of our results is checked.

\begin{figure}
\narrowtext
\centerline{
\epsfxsize=7.5truecm
\epsffile{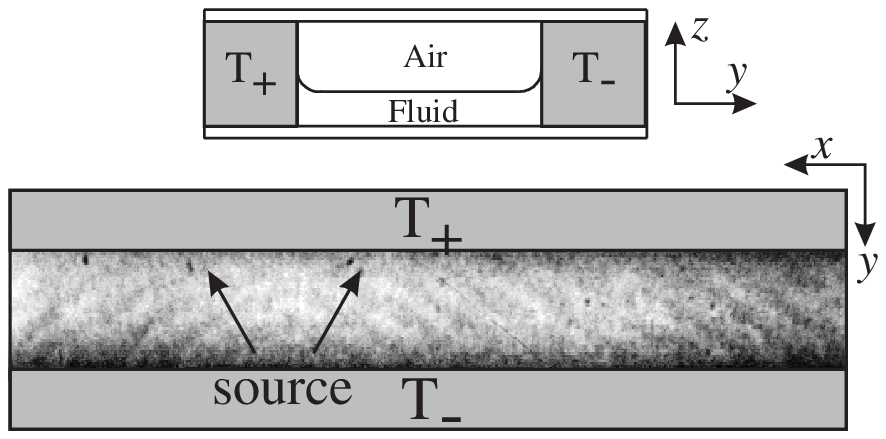}
}
\caption{Experimental cell and basic wave pattern.
Above: schematic side view. 
Below: top view and instantaneous shadowgraphic image of the wave pattern. 
}
\label{f1}
\end{figure}

The experimental setup is schematically described in Fig.~\ref{f1}. 
A layer of fluid (silicon oil of viscosity $\nu=0.65 {\rm cSt}$ 
and Prandtl number
$P=10$) of height $h$ is confined between two copper blocks maintained at
fixed temperatures $T_+$ and $T_-$ by thermostated water circulation, and a 
bottom glass plate. This forms a straight, narrow channel of 
length $L_x=25 {\rm cm}$ and width $L_y=2 {\rm cm}$. As soon as 
the temperature difference $\Delta T= T_+-T_-$ is not zero, a basic flow
sets in. It consists of a surface flow towards the cold side with a bottom 
recirculation.
Increasing $\Delta T$, the basic flow becomes unstable to traveling
hydrothermal waves~\cite{HYDWAVE} via a supercritical Hopf bifurcation \cite{HOPF}.
We observe these waves by low-contrast shadowgraphy, which captures the
 vertical average of the temperature gradient variations (surface waves exist,
but their effect is negligible). 
In this geometry,
the waves propagate away from a ``source region'' located arbitrarily on the 
cold wall, and the end boundaries at $x=0,L_x$ act as sinks with no apparent 
reflection (Fig.~\ref{f1}). 
For $h=1.2 {\rm mm}$, corresponding to
the experiment reported below, the source region emits
curved waves which become planar further away (Fig.~\ref{f1}, bottom) 
and propagate along the $x$-axis.

\begin{figure}
\narrowtext
\centerline{
\epsfxsize=8truecm
\epsffile{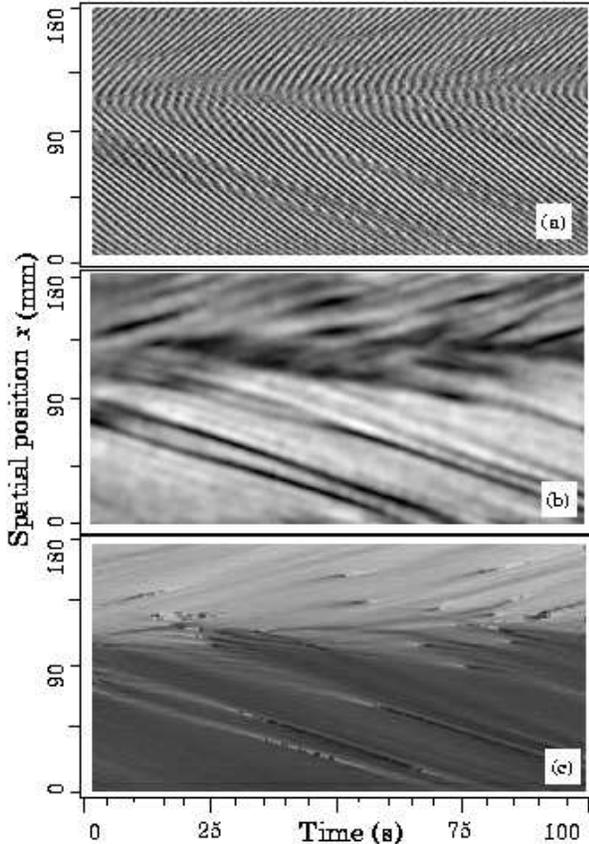}
}
\caption{Spatiotemporal evolution of the wave pattern at
$h=1.2 {\rm mm}$ and $\Delta T = 5.1 {\rm K}$ (yielding a Marangoni
number ${\rm Ma}\simeq 950$). 
Only a central portion of length 18.5cm is shown during 100s. 
(a): Original data.
(b): Evolution of the modulus $|A|$ (black: $|A|=0$; white: $|A|=1$) 
and (c): phase gradient $k$ (dark: $ k<0$; bright: $k>0$).}
\label{f2}
\end{figure}

Figure~\ref{f2}a presents a typical spatiotemporal evolution
as obtained from the acquisition,  with a 
fixed-gain camera, of a single 512-pixel line (of negligible width)
along the $x$-axis in the center of the cell. 
Here, the source
appears as a rather ill-defined, erratic object. (Closer to the Hopf
bifurcation, steady, regular, evolution is observed.)
Fourier analysis of diagrams such as Fig.~\ref{f2}a reveals that on
each side of the source only waves propagating away from the source are
present and that they are approximately monochromatic
(the second harmonic is two orders of magnitude smaller). 
More precisely, 
restricting ourselves to one side of the source (say $x\ge 90$ on 
Fig.~\ref{f2}a), we can write the recorded physical variable:
\begin{equation}
V(x,t)=A(x,t) \exp [ i ( k_0 x - \omega_0 t)] \;+\; {\rm c.c.}
\end{equation} 
where $k_0$ is the dominant wavelength, $\omega_0$ the basic frequency,
and $A$ a one-dimensional complex field describing the (large-scale)
modulations of this wave. (On the other side of the source, one has to change 
the sign of $k_0$.)

Using complex demodulation techniques, $A$ can be extracted from the 
experimental data. 
Figs.~\ref{f2}b,c show the spacetime evolution of 
$|A|$ and $k=q\pm k_0$  where $q=\partial_x \arg (A)$. In these pictures,
the localized deformations of the
waves visible in Fig.~\ref{f2}a clearly appear as propagating amplitude holes
across which the phase gradient varies rapidly. At some space-time points,
$|A|$ even vanishes and the phase gradient diverges: a space-time dislocation
occurs (Fig.~\ref{f2}a).
The amplitude holes can be seen as the objects mediating the evolution
to wave patterns more regular than those emitted by the source.

Our system clearly calls for a one-dimensional model.
The waves arise via a supercritical Hopf bifurcation. 
Away from the source, they propagate only in one direction.
All this indicate that
the evolution of $A$ could be governed by a single CGLE on each side of the
source, even though the regime studied here 
takes place at finite distance from 
threshold (for $h=1.2{\rm mm}$, $\Delta T_{\rm c} \simeq 4.3 {\rm K}$, and thus the 
relative distance to threshold is $\varepsilon = 0.19$ for 
$\Delta T = 5.1 {\rm K}$).
 We thus suppose that $A$ obeys:
\begin{eqnarray}
\tau_0 (\partial_t + v_{\rm g}\partial_x)A  = &  \nonumber \\
  \varepsilon A 
 + & \xi_0^2 (1\!+\!i \alpha)\partial_{xx} A  - g (1\!+\!i\beta)|A|^2 A
\label{cgle}
\end{eqnarray}
where $v_{\rm g}$ is the group velocity of the waves, $\tau_0$ and
$\xi_0$ are the basic time- and length-scales of the wave modulations, 
and $g$ is a real number. Below, we estimate, from the data
of Fig.~\ref{f2}, all the coefficients of Eq.~(\ref{cgle}) and check the
overall consistency of our hypothesis.

\begin{figure}
\centerline{
\epsfxsize=4truecm
\epsffile{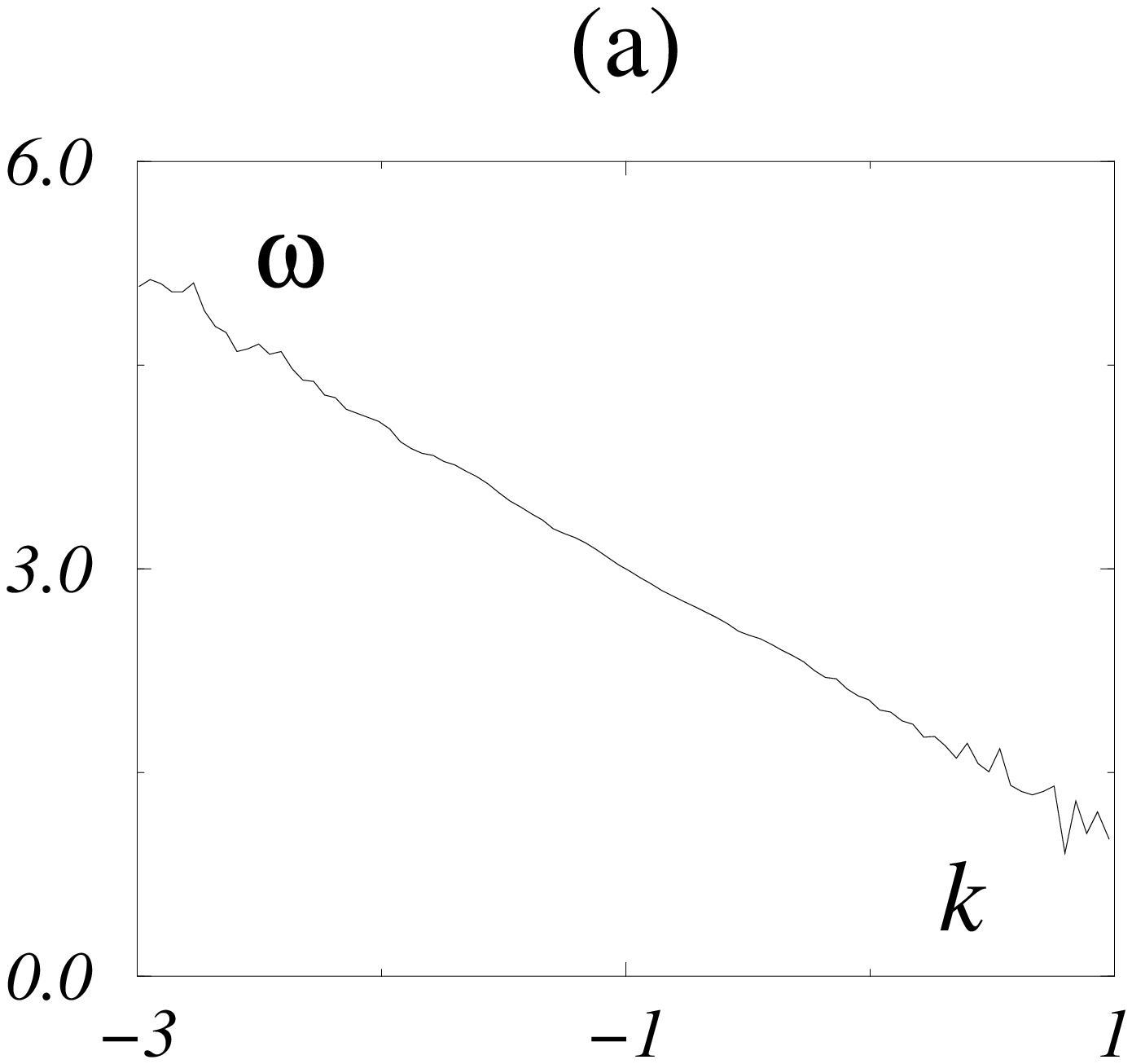}
\hspace{0.2cm}
\epsfxsize=4truecm
\epsffile{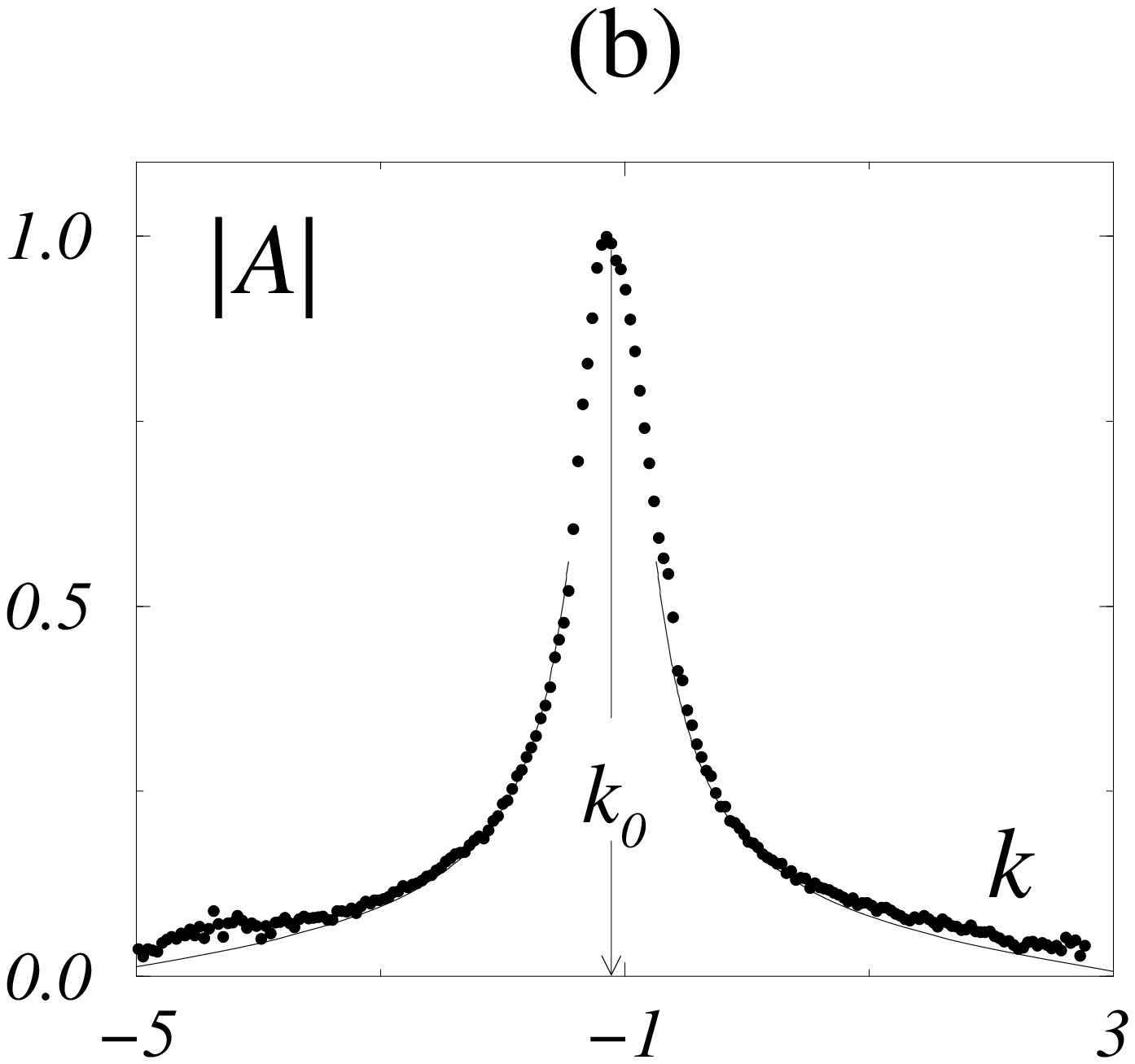}
}
\vspace{0.2cm}
\centerline{
\epsfxsize=4truecm
\epsffile{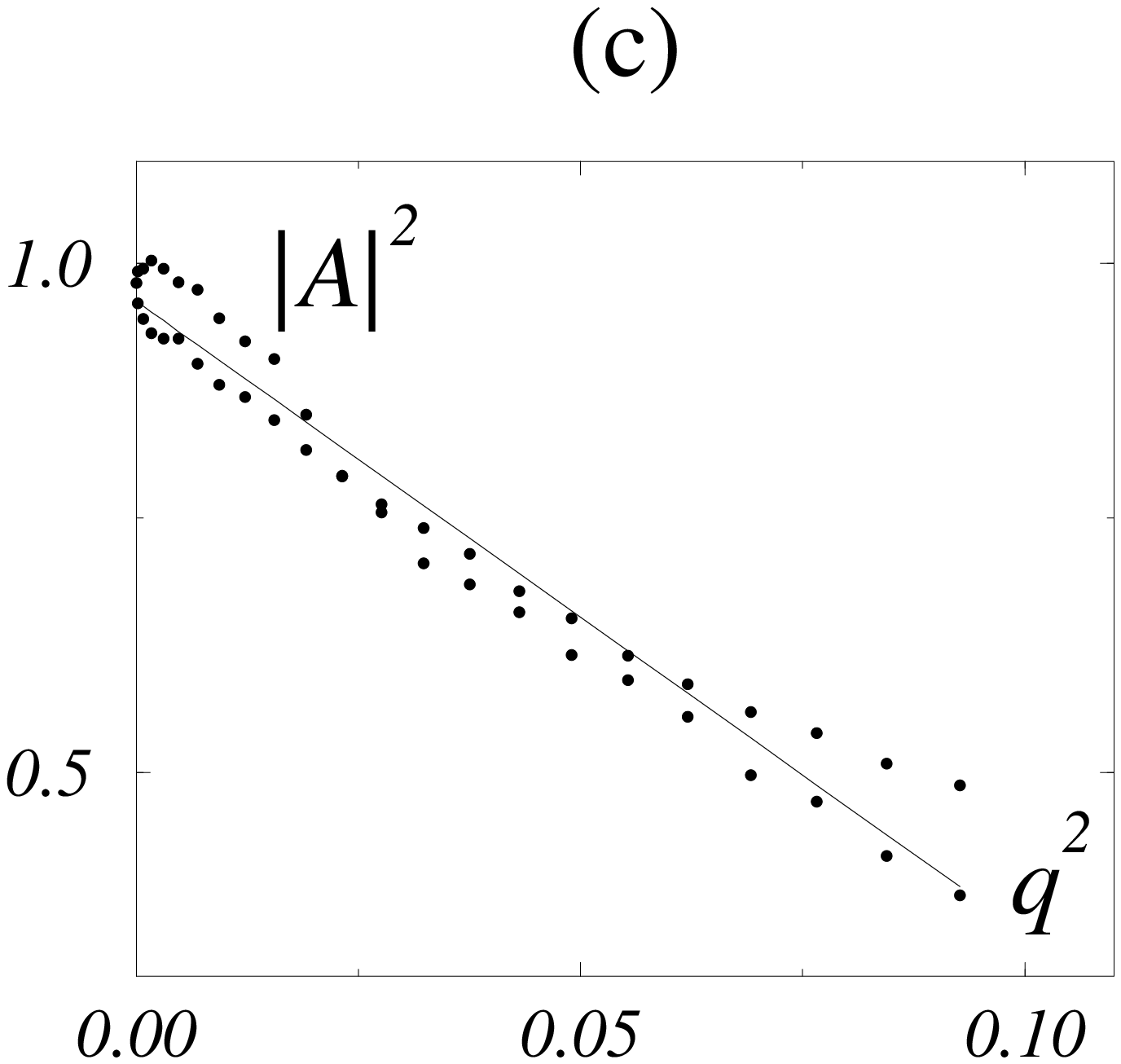}
\hspace{0.2cm}
\epsfxsize=4truecm
\epsffile{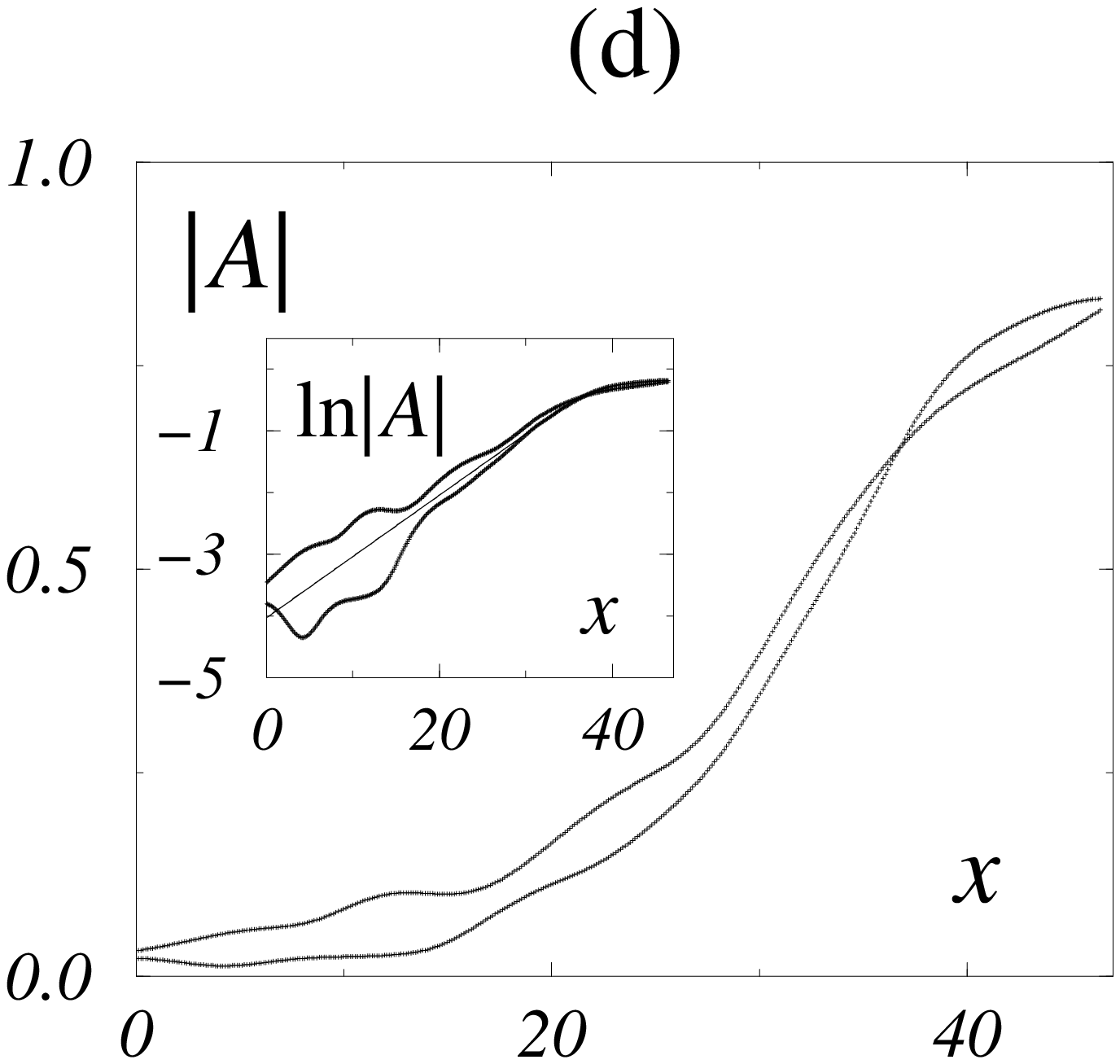}
}
\vspace{0.2cm}
\caption{Analysis of the data of Fig.~\protect\ref{f2} situated at the left
of the source ($x\le 90{\rm mm}$).
(a-c): for each bin around 
a given $k$ value, all the corresponding space-time 
points were determined, and averages calculated on each bin.
(a): local frequency $\omega$ vs local wavenumber $k$; the slope of a linear
fit gives the group velocity $v_{\rm g}$.
(b): local amplitude $|A|$ vs $k$; solid line: see text.
(c): $|A|^2$ vs $q^2$ (same data as in (b)).
(d): two instantaneous profiles of $|A|(x)$ near the source taken at time
0 and 30s; insert: $\ln |A|$
vs $x$ and linear fit (thin line).}
\label{f3}
\end{figure}

The linear part of the variation of the local  frequency $\omega$ 
with the local wavenumber $k$ yields our estimate of the group velocity:
$v_{\rm g}=\partial\omega/\partial k \simeq -1.16 {\rm mm/s}$ 
(Fig.~\ref{f3}a). This is consistent with the average value of the velocity 
of small perturbations, estimated at $-1.15\pm 0.25 {\rm mm/s}$, 
to be compared to the phase velocity 
$v_\phi=\omega_0/k_0 \simeq -2.8 {\rm mm/s}$.
This confirms that the source is indeed a source, since perturbations
do propagate outward.

Fig.~\ref{f3}b shows the variation of $|A|$ with $k$ as determined from 
the portion of Fig.~\ref{f2} at the left of the source ($x\le 90 {\rm mm}$). 
The maximum amplitude
is observed for the basic wavenumber: $k_0\simeq -1.11 {\rm mm}^{-1}$. 
Space-time points away from the localized amplitude holes correspond to
the large~$|A|$ (say $|A|>0.5$) portion of the curve. Locally
around these points, the solution of (\ref{cgle}) is expected
to be close to one of the phase-winding solutions
of wavevector $q=k+k_0$  (see e.g. \cite{MCCPCH}):
\begin{eqnarray}
A&=&A_q \exp [i (q x - \omega_q t)] \;\;\;{\rm with}\;\;\;
A_q^2= (\varepsilon - \xi_0^2 q^2)/g \nonumber \\ 
& & {\rm and}\;\;\;
\omega_q = [\varepsilon\beta+(\alpha-\beta)\xi_0^2 q^2]/\tau_0 -v_{\rm g}q
\label{wave}
\end{eqnarray}
The linear variation of $|A|^2$ with $q^2$ is confirmed
in Fig.~\ref{f3}c, yielding $\xi_0/\sqrt{\varepsilon}\simeq 2.53 {\rm mm}$
and $\sqrt{\varepsilon/g} \simeq 0.00054$ (a.u.). Note that we thus have 
$L_x \gg \xi_0/\sqrt{\varepsilon} \approx 3/k_0$: the cell is effectively 
``infinite'' and the variations of $A$ occur on scales 
significantly larger than the basic length $k_0^{-1}$.

Timescale $\tau_0$ can be estimated from the real part of the spatial
linear growth rate of waves near the source, which is equal to
$\varepsilon/(\tau_0 v_{\rm g})$ \cite{CROQ}.
From Fig.~\ref{f3}d, we find $\tau_0/\varepsilon \simeq 8.5 \pm 0.5 {\rm s}$,
about four times the basic period $2\pi/\omega_0 \simeq 2.03 {\rm s}$,
confirming that the variations of $A$ are slow compared to the basic 
oscillations. Note that $\tau_0$ is of the same order as the viscous
diffusion time $h^2/\nu =2.2 {\rm s}$.

At this stage, all the basic scales of Eq.~(\ref{cgle}) have been
estimated. To determine the remaining two parameters $\alpha$ and $\beta$,
global quantities deriving from the ``wave part'' of the data could, 
in principle, be sufficient. For example,
Fig.~\ref{f3}a could be used to extract the expected
variation of $\omega_q$ with $q$. But the data is too
noisy to yield any meaningful estimate of  $\alpha$ and $\beta$.
Moreover, as long as the source is not controlled, the ``input'' waves
cannot be varied at will to explore the family of solutions (\ref{wave}), 
contrary to other experimental situations \cite{CROQ}.
We now focus instead on the localized amplitude holes already mentioned.

Many localized, propagating objects connecting two phase-winding solutions
have been observed numerically in the one-dimensional CGLE \cite{NONLIN}.
Analytical methods are largely limited, so far, to solutions depending only on
the reduced variable $\xi=x-v_{\rm h} t$ where $v_{\rm h}$ is the (constant)
velocity of the object \cite{WVSPCH,CONTE}. 
Using this Ansatz, the CGLE reduces to a third-order
ordinary differential equation (ODE) whose fixed points are 
the phase-winding solutions (\ref{wave}).
Localized objects connecting
two such solutions of wavevector $q_{\rm L}$ and $q_{\rm R}$ appear, within
the Ansatz, as homoclinic ($q_{\rm L}=q_{\rm R}$) or heteroclinic 
($q_{\rm L}\ne q_{\rm R}$) orbits.

\begin{figure}
\centerline{
\epsfxsize=4truecm
\epsffile{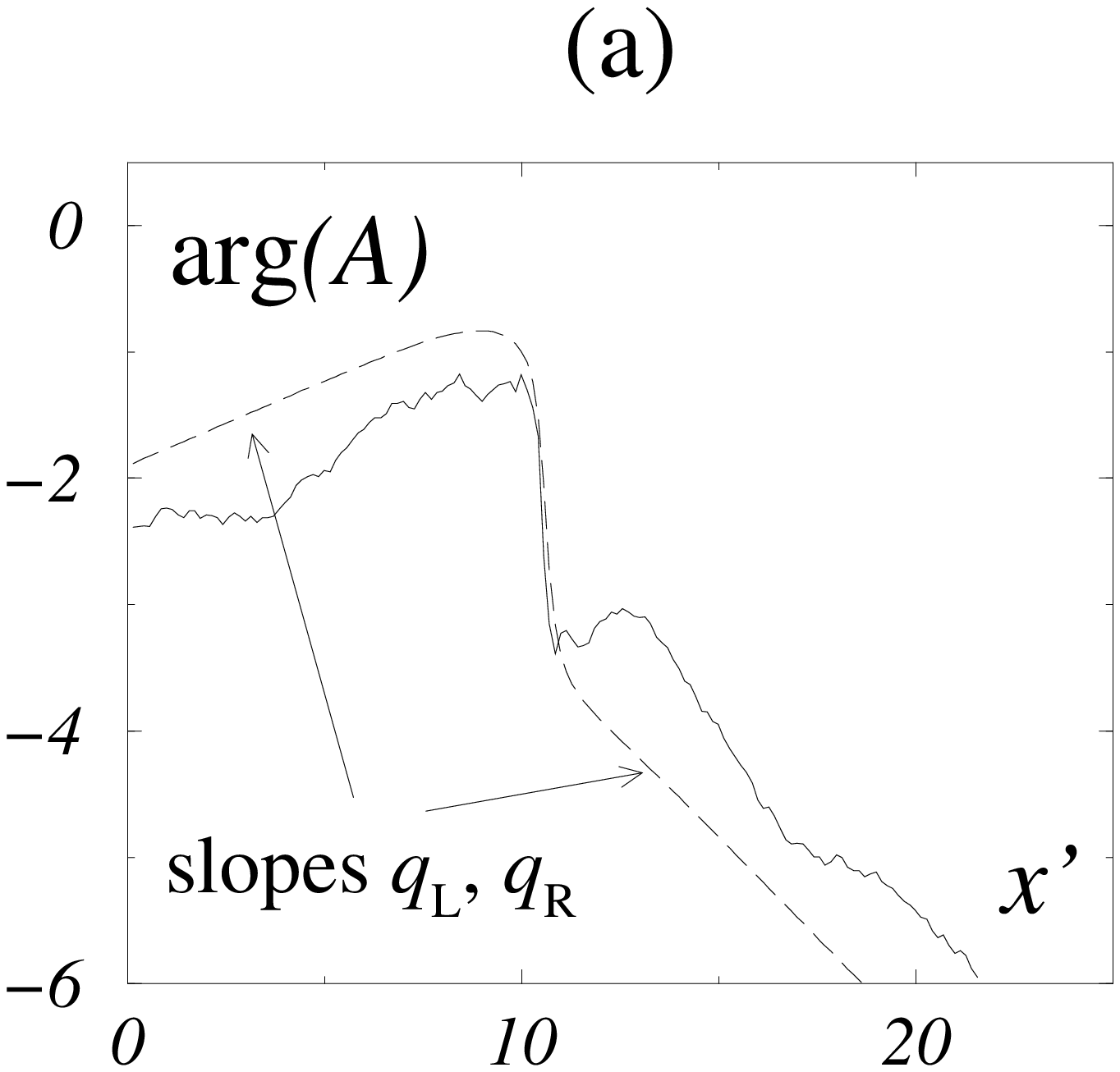}
\epsfxsize=4truecm
\epsffile{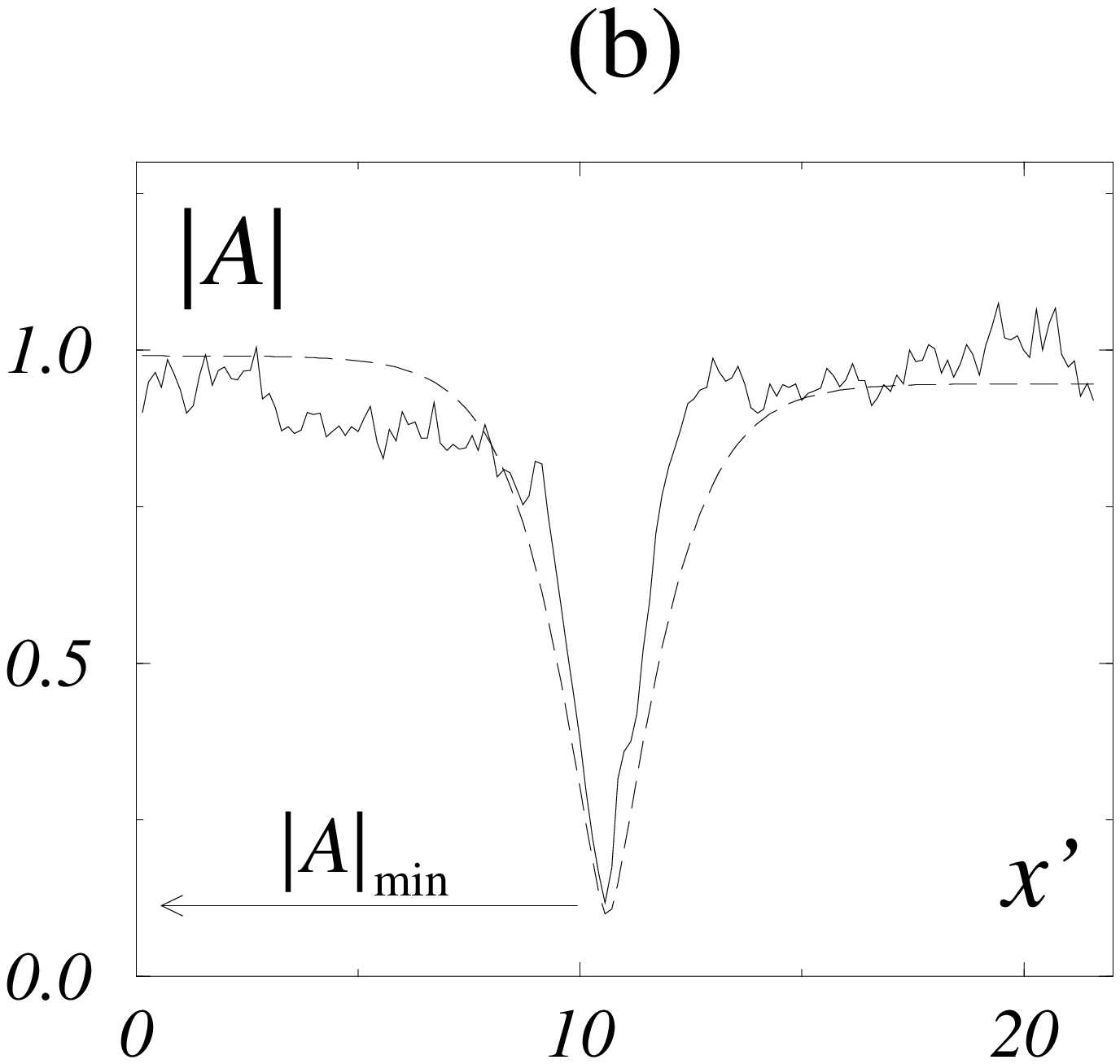}
}
\vspace{0.2cm}
\caption{Amplitude hole. 
Solid lines: experimental data  extracted from a cut of
Fig.~\protect\ref{f2} at $t=60 {\rm s}$. Dashed lines: 
corresponding Bekki-Nozaki hole solution of the equivalent CGLE.
(rescaled variable $x'=x \protect\sqrt{\varepsilon} / \xi_0$).}
\label{f4}
\end{figure}

The holes observed in Fig.~\ref{f2} are not stable structures
connecting two infinite phase winding solutions, but they subsist 
long enough, and can be sufficiently isolated to reveal that their wings
are indeed well described by phase winding solutions. (As a matter of fact,
we repeatedly assimilated, above, the large-amplitude regions separating
the holes to portions of these solutions.)
Fig.~\ref{f4} shows an isolated hole extracted from Fig.~\ref{f2}. 
One can measure rather accurately the two wavenumbers 
$q_{\rm L}=0.13$ and $q_{\rm R}=-0.32$ (in the CGLE frame) connected	
by the central hole, which can thus be tentatively seen as an heteroclinic
orbit in the ODE Ansatz. 

All sufficiently localized structures on
Fig.~\ref{f2} also connect two different wavenumbers. 
This rules out the homoclinic holes recently studied
by van Hecke \cite{HECKE}, and leave, as possible candidates, the family of
hole solutions found by Bekki and Nozaki \cite{NB}. The explicit form of these
solutions is too lengthy to be given here (see, e.g., \cite{CONTE}). 
They form a one-parameter family (at fixed $\alpha$, $\beta$) 
which can be parametrized by, e.g., the velocity $v_{\rm h}$ of the 
hole. They take the shape of an exponentially-localized amplitude hole
with a minimum amplitude $|A|_{\min}$ accompanied
by a rapid phase shift $\sigma$. 

\begin{figure}
\centerline{
\epsfysize=5truecm
\epsffile{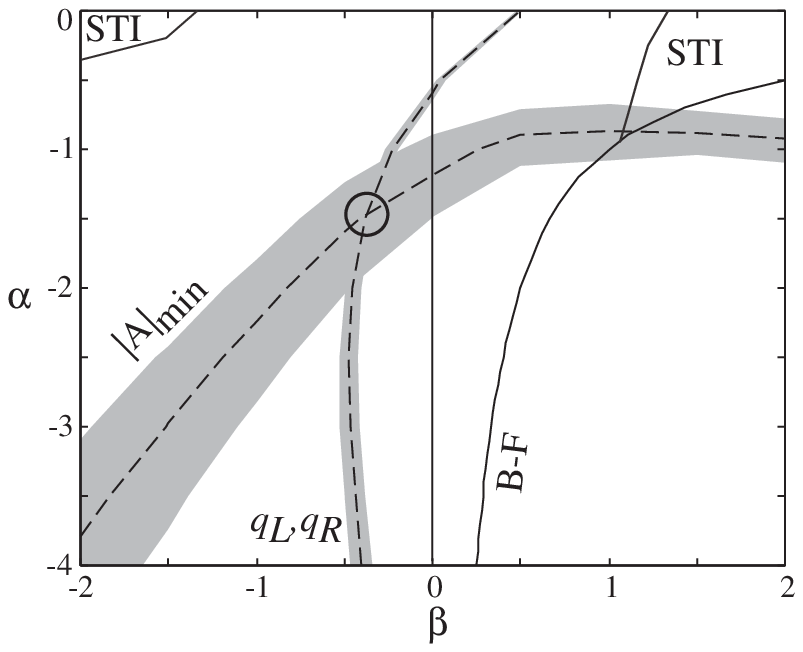}
}
\caption{Parameter plane of the CGLE. STI: spatiotemporal intermittency 
regions (from \protect\cite{NONLIN}). B-F: Benjamin-Feir line below which
all solutions (\protect\ref{wave}) are unstable. Dashed lines: see text
(grey areas: error bars). Circle: estimated $(\alpha,\beta)$ values.}
\label{f5}
\end{figure}

To compare data such as that of Fig.~\ref{f4} to these solutions,
we need to determine the values of $\alpha$ and $\beta$ and the ``optimal''
solution of the corresponding family. We proceed as follows:
we estimate $q_{\rm L}$,  $q_{\rm R}$, and $|A|_{\min}$ 
from the data since we found these were the characteristics of the hole 
for which the most accurate measurement can be made.
We find, for all values of the $(\alpha,\beta)$ plane where it exists,
the Bekki-Nozaki hole solution with the measured value of 
$q_{\rm L}+q_{\rm R}$. We then select the (codimension 1) subsets of the
 $(\alpha,\beta)$ plane where, moreover, this hole solution possesses the
measured value of $q_{\rm L}$ or the estimated value of $|A|_{\min}$
(Fig.~\ref{f5}, dashed lines). These two lines intersect, yielding the 
desired values of $\alpha$ and $\beta$. Taking into account the error bars on 
 $q_{\rm L}$,  $q_{\rm R}$, and $|A|_{\min}$, we find $\alpha= -1.5 \pm 0.5$
and $\beta = -0.4\pm 0.05$. By the same token, the hole solution is also 
uniquely determined (Fig.~\ref{f4}, dashed lines). Its velocity $v_{\rm h}$,
its width, its phase shift $\sigma$ are all found consistent with
the  data.

We repeated the procedure for other amplitude holes found in Fig.~\ref{f2}. 
We found almost the same $\alpha$ and $\beta$ values, to the accuracy of the
estimates. This strengthens the confidence in the results, since different
objects moving at different velocities, connecting different wavenumbers
all yield the same parameter values.
We also performed a final check, by plotting, for the estimated values
of $\alpha$ and $\beta$, the variation of $|A|_{\min}$ with the local phase
gradient at the ``bottom'' of the hole 
along the family of solutions (Fig.~\ref{f3}b,
solid line). The agreement with the small-$|A|$ values measured on 
Fig.~\ref{f2} is very good. This is an additional indication that all the
low-amplitude points are indeed located ``inside'' Bekki-Nozaki holes.

For the estimated values of $\alpha$ and $\beta$, the CGLE is in the
parameter region where the phase-winding solutions (\ref{wave})
are linearly stable for $|q|$ small enough and no
 sustained spatiotemporal disorder exists in one space dimension
\cite{NONLIN}. Moreover, the
Bekki-Nozaki amplitude holes are linearly unstable \cite{NBSTAB}. 
This is not in contradiction with the dynamics observed in the
experiment: the Bekki-Nozaki holes, although unstable, exist, and
can constitute important building blocks of even chaotic dynamics
\cite{NONLIN}. The 
waves emitted by the source can be locally attracted to this family of
unstable fixed points before escaping along its unstable manifold
(a mechanism also invoked by van Hecke in \cite{HECKE}). The tendency
of the waves trains to become more regular away from the source 
(see Fig.~\ref{f2}) is consistent with disorder being only transient in
the CGLE with the estimated parameter values.

In summary, we have presented experimental results on the dynamics of the
nonlinear hydrothermal waves traveling in a laterally-heated fluid layer.
We have shown that,
although the regime studied here is rather far from the onset of waves,
the large-scale modulations
of the basic pattern are governed  by a one-dimensional complex Ginzburg-Landau
equation and we estimated its full set of coefficients. This was made 
possible by showing that
the localized amplitude holes observed experimentally correspond to
the Bekki-Nozaki hole solutions of the CGLE. The overall consistency of
our results was checked. Since the operating regime of the CGLE at
the  estimated  parameter values does not exhibit sustained disorder, it
would be interesting to analyze experimental data collected at other 
parameter values in the hope of reaching spatiotemporal chaos regimes
of the type exhibited by the CGLE. This is left for future work, together
with an attempt to obtain a better control of the system by forcing 
the behavior of the source.
More generally, we believe that using
the localized structures or defects of pattern-forming systems to
determine quantitatively their relevant amplitude equations can be
a rewarding approach to this difficult experimental problem.

J.B. thanks the spanish government for support 
through project PB95-0578A (DGICYT) and a postdoctoral grant
(SEUI, Mi\-nis\-terio de Educaci\'on y Ciencia).

\end{multicols}

\end{document}